\documentstyle[12pt,epsf]{article}
\def\epsfig\#1{}
\newlength{\dinwidth}
\newlength{\dinmargin}
\begin{document}
\newcommand{\GeVsq}     {\mbox{\rm ~GeV}^2}
\newcommand{\sleq} {\raisebox{-.6ex}{${\textstyle\stackrel{<}{\sim}}$}}
\newcommand{\sgeq} {\raisebox{-.6ex}{${\textstyle\stackrel{>}{\sim}}$}}
\newcommand{\ptmiss}{\mbox{$\not\hspace{-1mm}{P}_t$$}}

\title{
{ \bf Measurement of Charged and Neutral Current $e^-p$
Deep Inelastic Scattering Cross Sections at High $Q^2$  }
\\
\author
{\rm ZEUS Collaboration\\}
                                            }
\date{ }
\maketitle
\vspace{5 cm}
\begin{abstract}
\noindent
Deep inelastic $e^-p$ scattering has been studied in both the
charged-current (CC) and neutral-current (NC) reactions at momentum
transfers squared, $Q^2$, between 400 GeV$^2$ and the kinematic limit of
87500 GeV$^2$ using the ZEUS detector at the HERA $ep$ collider. The CC
and NC total cross sections, the NC to CC cross section ratio, and the
differential cross sections, $ d\sigma/dQ^2 $, are presented.  For $Q^2
\simeq M_W^2$, where $M_W$ is the mass of the $W$ boson, the CC and NC
cross sections have comparable magnitudes, demonstrating the equal
strengths of the weak and electromagnetic interactions at high $Q^2$. The
$Q^2$ dependence of the CC cross section determines the mass term in the
CC propagator to be $M_{W} = 76 \pm 16 \pm 13$~GeV.
\end{abstract}

\vspace{-20cm}
\begin{flushleft}
\tt DESY 95-053 \\
March 1995 \\
\end{flushleft}

\setcounter{page}{0}
\thispagestyle{empty}

\newpage
\def\3{\ss}
\parindent 0cm
\footnotesize
\renewcommand{\thepage}{\Roman{page}}
\begin{center}
\begin{large}
The ZEUS Collaboration
\end{large}
\end{center}
M.~Derrick, D.~Krakauer, S.~Magill, D.~Mikunas, B.~Musgrave,
J.~Repond, R.~Stanek, R.L.~Talaga, H.~Zhang \\
{\it Argonne National Laboratory, Argonne, IL, USA}~$^{p}$\\[6pt]
R.~Ayad$^1$, G.~Bari, M.~Basile,
L.~Bellagamba, D.~Boscherini, A.~Bruni, G.~Bruni, P.~Bruni, G.~Cara
Romeo, G.~Castellini$^{2}$, M.~Chiarini,
L.~Cifarelli$^{3}$, F.~Cindolo, A.~Contin, M.~Corradi,
I.~Gialas$^{4}$,
P.~Giusti, G.~Iacobucci, G.~Laurenti, G.~Levi, A.~Margotti,
T.~Massam, R.~Nania, C.~Nemoz, \\
F.~Palmonari, A.~Polini, G.~Sartorelli, R.~Timellini, Y.~Zamora
Garcia$^{1}$,
A.~Zichichi \\
{\it University and INFN Bologna, Bologna, Italy}~$^{f}$
% ...
\\[6pt]
\clearpage
\hspace*{1mm}
$^{ 1}$ supported by Worldlab, Lausanne, Switzerland \\
\hspace*{1mm}
$^{ 2}$ also at IROE Florence, Italy  \\
\hspace*{1mm}
$^{ 3}$ now at Univ. of Salerno and INFN Napoli, Italy  \\
\hspace*{1mm}
$^{ 4}$ supported by EU HCM contract ERB-CHRX-CT93-0376 \\
\hspace*{1mm}
% ...
$^{16}$ now with OPAL Collaboration, Faculty of Physics at Univ. of
        Freiburg \\
% ...

\begin{tabular}{lp{15cm}}
$^{a}$ &supported by the Natural Sciences and Engineering Research
         Council of Canada (NSERC) \\
$^{b}$ &supported by the FCAR of Qu\'ebec, Canada\\
$^{c}$ &supported by the German Federal Ministry for Research and
         Technology (BMFT)\\
$^{d}$ &supported by the MINERVA Gesellschaft f\"ur Forschung GmbH,
         and by the Israel Academy of Science \\
$^{e}$ &supported by the German Israeli Foundation, and
         by the Israel Academy of Science \\
$^{f}$ &supported by the Italian National Institute for Nuclear Physics
         (INFN) \\
$^{g}$ &supported by the Japanese Ministry of Education, Science and
         Culture (the Monbusho)
         and its grants for Scientific Research\\
$^{h}$ &supported by the Korean Ministry of Education and Korea Science
         and Engineering Foundation \\
$^{i}$ &supported by the Netherlands Foundation for Research on Matter
         (FOM)\\
$^{j}$ &supported by the Polish State Committee for Scientific Research
         (grant No. SPB/P3/202/93) and the Foundation for Polish-
         German Collaboration (proj. No. 506/92) \\
$^{k}$ &supported by the Polish State Committee for Scientific
         Research (grant No. PB 861/2/91 and No. 2 2372 9102,
         grant No. PB 2 2376 9102 and No. PB 2 0092 9101) \\
$^{l}$ &partially supported by the German Federal Ministry for
         Research and Technology (BMFT) \\
$^{m}$ &supported by the German Federal Ministry for Research and
         Technology (BMFT), the Volkswagen Foundation, and the Deutsche
         Forschungsgemeinschaft \\
$^{n}$ &supported by the Spanish Ministry of Education and Science
         through funds provided by CICYT \\
$^{o}$ &supported by the Particle Physics and Astronomy Research
        Council \\
$^{p}$ &supported by the US Department of Energy \\
$^{q}$ &supported by the US National Science Foundation
\end{tabular}

\newpage
\pagenumbering{arabic}
\setcounter{page}{1}
\normalsize

\section{Introduction}
\label{s:intro}
Lepton-nucleon scattering is an important technique for
studying the constituents of the nucleon and
their interactions.
In the Standard Model\cite{PDG1}, electron-proton ($ep$) scattering occurs
via the
exchange of gauge bosons ($\gamma$, $Z^0$, $W^\pm$).
At long wavelengths (small momentum transfers),
interactions of the massless photon dominate over the
exchange of the heavy gauge bosons. However, at the $ep$ storage ring HERA,
for the first time, scattering can be observed at
sufficiently short wavelengths (large momentum transfers) that
the `weak' and `electromagnetic' scattering cross sections have comparable
magnitudes.

Neglecting longitudinal  structure functions and radiative corrections,
the differential cross section for deep inelastic
scattering (DIS) with unpolarized $e^-p$ beams
can be expressed as\cite{Ruckl}:

$$ \frac{d^2 \sigma}
{dx dQ^2} = \frac{2\pi \alpha^2}{x Q^4}
\left[ \{ 1+(1-y)^2 \}
 {\cal F}_2 + \{ 1-(1-y)^2 \} x {\cal F}_3 \right]
$$

\noindent{where the ${\cal F}_i(x,Q^2)$
functions  describe the proton structure and couplings.
In this equation, $Q^2$ is the negative square of the four-momentum transfer,
$y$ is the fractional
energy transfer from the lepton in the proton rest frame,
$\alpha$ is the electromagnetic fine-structure constant and
 $x$ in the quark-parton model is the momentum fraction of the proton
carried by the quark struck by the exchanged boson.
These variables are related by
$Q^2 = s x y $, where  $\sqrt{s}$ is the center-of-mass energy.
 The
${\cal F}_i$ can be expressed as  sums over quark flavors, $f$,
of the quark densities inside the proton, $q_f(x,Q^2)$,
weighted according to the
gauge structure of the scattering amplitudes. For the
neutral-current (NC) reaction, $e^- p \rightarrow e^- X$,
mediated by $\gamma $ and $Z^{0}$ exchange, they can be
written as:}

\begin{eqnarray*}
 {{\cal F}_2^{NC}} = \sum_f q_f^+
\left[ e_f^2 +  2 v_e v_f e_f {{\cal P}_Z} +
 (v_e^2+a_e^2)(v_f^2+a_f^2)  {{\cal P}^2_{Z} } \right] \\
\end{eqnarray*}

\begin{eqnarray*}
 x{{\cal F}_3^{NC}} = \sum_f q_f^-
\left[ -2 a_e a_f e_f{{\cal P}_Z} +
  \left( 4v_e a_e v_f a_f \right) {{\cal P}^2_Z} \right]
\end{eqnarray*}

\noindent{ where $q_f^{\pm}= \{ xq_f(x,Q^2)\pm x\bar{q}_f(x,Q^2) \} $,
$a_e$ and $v_e$ are the axial- and vector-couplings
of the $e^-$ to the $Z^0$, and
$a_f$ and $v_f$ are the analogous couplings for
a quark of flavor $f$ which has electric charge $e_f$\cite{PDG1}.
${\cal P}_Z$ is the ratio of $Z^0$-to-photon propagators, given by
${\cal P}_Z$ ~=~$Q^2/(Q^2 + M_Z^2)$, where $M_Z$ is the mass of the
$Z$ boson. }

For charged-current (CC) scattering, $e^- p \rightarrow \nu_e X$, in which
$W^{\pm}$ bosons are exchanged, the functions are:
$$
{\cal F}_2^{CC}
 = \frac{{x{\cal P}^2_W }}{8 \sin^4\theta_W}
 \sum_{k,m} \left[ |V_{km}|^2 u_k + |V_{mk}|^2
 \bar{d}_m \right]
$$
\
$$
x{\cal F}_3^{CC}
 = \frac{{x{\cal P}^2_W }}{8 \sin^4\theta_W}
 \sum_{k,m} \left[ |V_{km}|^2 u_k - |V_{mk}|^2
 \bar{d}_m \right]
$$
where $k$ and $m$ are the generation indices of
up-type  quarks, $u_k(x,Q^2)$, and down-type antiquarks, $\bar{d}_m(x,Q^2)$,
$V$ is the Cabibbo-Kobayashi-Maskawa quark mixing matrix, $\theta_W$ is
the weak mixing angle, and
${\cal P}_{W} ={Q^2}/(Q^2+M^2_W)$.
 At lowest order, $G_F M^2_W = \pi \alpha /
\sqrt{2} \sin^2\theta_W$, where $G_F$ is the Fermi constant.

In 1993, HERA  collided 26.7 GeV $e^-$ with 820 GeV $p$, giving
$\sqrt{s}=296$ GeV.
Due to this high center-of-mass energy,
  DIS can be investigated at much higher $Q^2$ at HERA than in existing
fixed target experiments.
The predicted  DIS cross sections at fixed $x$ over a large $Q^2$ range depend
 both on the electroweak theory for  the
propagators and couplings and on Quantum Chromodynamics (QCD)
 for the parton density evolution. The structure functions, ${\cal F}_2$,
have been measured\cite{Feltesse} in $ep$ and $\mu p$ scattering up to
$Q^2 \sim 5 (150) \mbox{\rm ~GeV}^2 $ for $x = 0.03 (0.3)$. The parton density
distributions\cite{mrsd,cteq} inferred from those measurements were
extrapolated to our $Q^2$ region using the next-to-leading-order QCD evolution
equations\cite{GLAP}. At $x$
of 0.03 (0.3), the up-quark density is predicted to change by $21\% $
($-39\%$) as $Q^2$ increases from 5 $\mbox{\rm ~GeV}^2$ to 16000 $\mbox{\rm
{}~GeV}^2$. In contrast,
the NC propagator varies by 7 orders of magnitude over the same $Q^2$ interval.

 This paper reports measurements of integrated and differential
cross sections, $d\sigma/dQ^2$, for NC and CC
DIS with $Q^2 > 400 \mbox{\rm ~GeV}^2 $  using a
luminosity of $0.540\pm 0.016~ \mbox{\rm pb}^{-1}$.
ZEUS\cite{zeusf2} and H1\cite{h1f2} have previously reported on NC DIS
cross section measurements at lower $Q^2$.
The H1 experiment has also
measured the CC total cross section\cite{h1cc} and demonstrated
that at large $Q^2$ the mass  in the CC propagator is finite.

\section{The ZEUS detector and trigger}

ZEUS\cite{status} is a multipurpose, magnetic detector,
designed  especially to measure DIS.
Charged particles are tracked by drift chambers operating in
an axial magnetic field of 1.43~T.
The superconducting solenoid is
surrounded by a compensating uranium-scintillator
calorimeter (CAL)
with
an
electromagnetic (hadronic) energy resolution of 18\%$/\sqrt{E(\mbox{\rm GeV})}$
(35\%$/\sqrt{E(\mbox{\rm GeV})}$)  and a subnanosecond
time resolution. The CAL covers the angular range between $2.2^\circ$ and
$176.5^\circ$.
ZEUS used a right-handed coordinate system, centered at the nominal interaction
point ($z=0$), defined with positive $z$ along the direction of the proton
beam and positive $y$ upwards.
The CAL is segmented in depth into electromagnetic and hadronic
sections, with a total thickness of 4 to 7 interaction lengths.
Surrounding the CAL is an iron magnetic return yoke
instrumented for muon detection. For this analysis,
   the muon  detectors
were used to identify cosmic-ray induced triggers.
The luminosity is measured by the rate of high-energy photons from the
reaction
$ep \rightarrow ep\gamma$ detected in a
lead-scintillator calorimeter located  $z=-107~ \mbox{\rm m}$ from the
interaction region.

Data were collected with a three-level trigger.
The first-level trigger was based on electromagnetic energy, transverse
energy and total energy deposits in the CAL\cite{zeusf2}.
The thresholds, between 2 and
15 GeV, were well below the offline selection cuts.
The second-level trigger rejected $p$-gas events (proton interactions
with residual gas in the beam pipe upstream of the detector)
recognized by CAL energy deposited at times early
relative to that of the $ep$ crossing.
The third-level trigger selected events  as
NC DIS candidates if $E-P_z$ exceeded 25 GeV, where $E$ and
$P_z$ are the summed energy and $z$-component of the momentum measured in the
calorimeter.  If no energy escapes through the rear
beam hole, $E-P_z\approx 2E_e$ where $E_e$ is the electron beam energy.
  Events were selected as CC DIS candidates if
\mbox{$\not\hspace{-1mm}{P}_t$}, the absolute value of the
missing  transverse momentum measured by the calorimeter,
exceeded 9 GeV, and there was either more than 10 GeV deposited in the forward
part of CAL or at least one track reconstructed in the drift chambers.

\section{Kinematic Reconstruction and Event Simulation}

As the ZEUS detector is nearly hermetic, it is possible
to reconstruct the kinematic variables $x$ and
$Q^2$ for NC DIS using different
combinations of the angles and energies of the scattered lepton and
hadronic system\cite{zeusf2}. Three methods were relevant to this analysis.
The electron ($e$) method uses  $E_e'$ and $\theta_e$,
the energy and polar angle of the scattered electron.
The hadronic, or Jacquet-Blondel (JB)\cite{JB}, method reconstructs $y$ and
$Q^2$
as $y_{JB}=(E_{h}-P_{z,h})/(2E_e)$ and
$Q^2_{JB}=P^2_{t,h}/(1-y_{JB})$, where $E_h$, $P_{z,h}$ and $P_{t,h}$ are
the energy, the $z-$component of momentum, and
the transverse momentum, of the hadronic
system.
The double angle (DA) method uses $\theta_e$ and
$\gamma_h$, the polar angle of the struck quark which is given by
$\cos\gamma_{h} = (P^2_{t,h} - (2 E_e y_{JB})^2)/
                  (P^2_{t,h} + (2 E_e y_{JB})^2)$.
The DA method, which  measures
$Q^2$ with  small bias and good
resolution,
was used to reconstruct NC events\cite{zeusf2}.
For CC DIS, the hadronic (JB) method was used.

The acceptances and measurement
resolutions for signal and background events were determined using
Monte Carlo methods. Simulated CC and NC DIS events, generated using
LEPTO\cite{lepto} interfaced to
HERACLES\cite{heracles} by DJANGO\cite{django}, were passed through
a GEANT\cite{geant} based detector simulation, and subsequently
analyzed with the same reconstruction and offline selection procedures
as the data. The calculated efficiencies and acceptances were found
to have negligible dependences on either the model of
the hadronic final state \cite{lepto,ariadne}
or the proton parton density parametrizations\cite{mrsd} used in the
simulation.

\section{NC selection and analysis}

\begin{figure}[tbp]
\epsfxsize=6in
\epsfysize=6.5in
\epsfbox[-90 40 517 607]{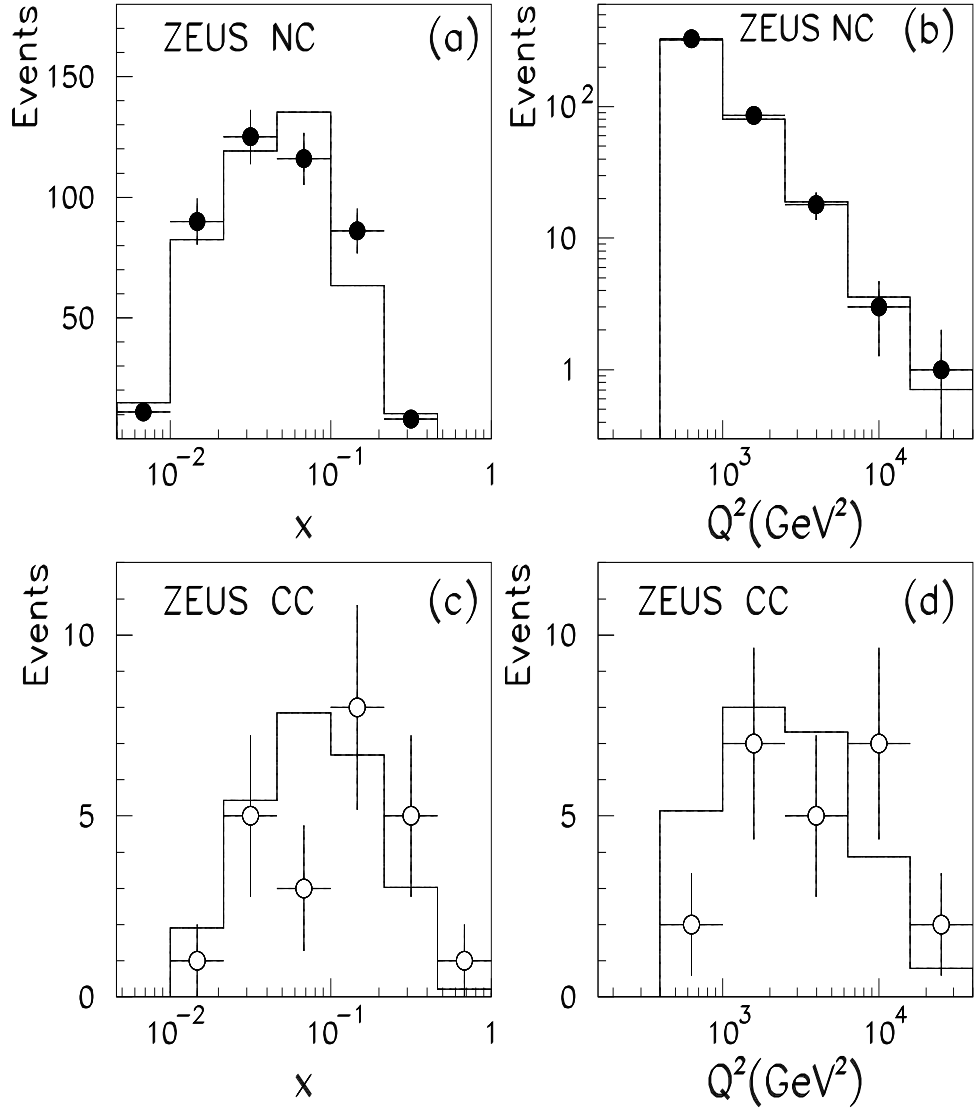}
\vspace*{22mm}
\caption[fig4]{\label{fig:dndx} (a) $x$ for NC events
(b) $Q^2$ for NC events (c) $x$ for CC events (d) $Q^2$ for CC events.
The points with error bars are ZEUS data. The histograms are the predicted
numbers of events from the absolutely normalized simulation.}
\end{figure}

The offline NC DIS event selection required an
electron candidate with $E_e' > 10$~GeV in the calorimeter
and  $E-P_z > 35$ GeV. To reject backgrounds from photoproduction events
with a fake electron (mostly $\pi^0$'s at
small polar angles) the electron candidate was required to
have a matching track and to satisfy $y_e < 0.95$.
Cosmic-ray triggers were rejected by requiring
$\mbox{$\not\hspace{-1mm}{P}_t$}/\sqrt{E_t} < 2$
GeV$^{\frac{1}{2}}$. A final cut
required $Q^2$, as reconstructed by the $DA$ and $e$ methods,
to be consistent: $0.7 < Q^2_e /Q^2_{DA} < 1.2 $.
After these selections, 436 events with $Q^2_{DA}>400$ GeV$^2$ remained.
The photoproduction background is less than 2\%. Over the full $y$ range
of $0 < y < 1 $, more than 85\% of all Monte Carlo NC DIS events with
$Q^2 > 400 \mbox{\rm ~GeV}^2 $
pass all of the above cuts.
The spectra of $x$ and $Q^2$ for the data and the Monte Carlo
simulation are shown in Figures \ref{fig:dndx}a,b.  The agreement is
satisfactory in both shape and
absolute magnitude.

The NC DIS cross sections in five bins of $Q^2$ between $ 400 \mbox{\rm ~GeV}^2
$ and the
kinematic limit at $87500 \mbox{\rm ~GeV}^2$ are given in Table
\ref{tab:dndbin}.
The cross section was calculated for each bin as $\sigma_{NC}
= (N_{NC} \cdot \delta r_{NC} )/({\cal L}\cdot{\cal A}_{NC}$) where
$N_{NC}$ is the number of NC DIS events
reconstructed in the bin, $\delta r_{NC}$ is the radiative
correction,  and $\cal{L}$ is the luminosity. The
acceptance for the bin, ${\cal A}_{NC}$, was calculated from
the NC DIS Monte Carlo event sample, as the ratio of the number
of events which pass all cuts and have the reconstructed $Q^2_{DA}$ in the
bin to the number of events with the true $Q^2$ in the bin.
${\cal A}_{NC}$  varies between 0.79 and 0.85.
HERACLES\cite{heracles} was used to calculate the radiative correction factor,
$\delta r_{NC}$, which was in the range  0.88 to 0.95 and  has been
applied to the data in order to obtain Born cross sections.

The systematic errors on ${\cal A}_{NC}$ include: a
4\% error assigned to the uncertainty of the calorimeter energy response; a
3\% uncertainty assigned to the efficiency of the calorimeter-track matching
for the electron; a
4\% error for the efficiency of the electron finding algorithm; and a
5\% error in the lowest $Q^2$ bin for the uncertainty
in the efficiency of the $Q^2_e/Q^2_{DA}$ cut.

\section{CC selection and analysis}

The CC DIS events are characterized by a large
\mbox{$\not\hspace{-1mm}{P}_t$}\ due to the final-state neutrino. The 36000
triggers
for this mode
were produced predominantly by
upstream
$p$-gas interactions or cosmic rays.
The offline CC DIS selection required  \mbox{$\not\hspace{-1mm}{P}_t$}\ $>
12$~GeV
and  a  vertex, formed from two or more tracks, within 45~cm of the nominal
interaction point.
Events with more than 40 tracks not
associated with the vertex were rejected.
To reduce the remaining $p$-gas background,
for which the reconstructed transverse energy was concentrated at small polar
angles, events with
$\mbox{$\not\hspace{-1mm}{P}_t$}^{outer}<0.7\mbox{$\not\hspace{-1mm}{P}_t$}$
were rejected, where
 \mbox{$\not\hspace{-1mm}{P}_t$}$^{outer}$ is
the missing transverse momentum in the calorimeter excluding
the $1.0\times 1.0$ m$^2$ region around the forward beam pipe.
The 117 candidates remaining were mostly cosmic-ray events, including
cosmic-ray muons coincident with a $p$-gas interaction. Single muons
were rejected on the basis  of their characteristic spatial distribution of
energy deposition in the calorimeter. Additionally, the times of all
energy deposits measured in the calorimeter were required to be
consistent with a single $ep$ interaction.
Events with track segments in three or more
muon chambers were also rejected.

The events passing all selection criteria were scanned and
one cosmic-ray event was removed, leaving
23 events with $Q^2 > 400 \mbox{\rm ~GeV}^2 $ in the final CC DIS sample.
{}From Monte Carlo simulations, we expect fewer than one
background event from photoproduction.

The hadronic energies in CC events were corrected for energy loss in material
between the vertex and the calorimeter with a multiplicative factor which
depended on the uncorrected $P_{t,h}$ and $E_h-P_{z,h}$. The
correction was determined using NC DIS events, for which
the hadronic four-momentum can be reconstructed using the DA method. The
correction factor varies between 1.03 (at small $E_h-P_{z,h}$) and
1.22 (at small $P_{t,h}$). Figures \ref{fig:dndx}c, d
show the reconstructed $x$ and $Q^2$
distributions for CC DIS sample with $Q^2>400$ GeV$^2$
compared to the Monte Carlo simulation. Within the limited statistics
of the data, the agreement is satisfactory.

The CC DIS cross sections, $\sigma_{CC}
=(N_{CC} \cdot \delta r_{CC}) /({\cal L}\cdot{\cal A}_{CC}$),
are shown in Table \ref{tab:dndbin}.
The acceptance, ${\cal A}_{CC}$,  is in the range 0.67 to 0.80, except
for the bin at largest $Q^2$ where it is $1.10$ due to  migration from
lower $Q^2$. Seventy-five percent  of Monte
Carlo CC DIS events generated with
$Q^2 > 400 \mbox{\rm ~GeV}^2$ pass all of the selection cuts.
The systematic errors on ${\cal A}_{CC}$ include: a
5\% estimated uncertainty from the dependence on the
\mbox{$\not\hspace{-1mm}{P}_t$}\ and
the $\mbox{$\not\hspace{-1mm}{P}_t$}^{outer}/
\mbox{$\not\hspace{-1mm}{P}_t$}$ thresholds; a
5\% assigned error on the efficiency to reconstruct a vertex with two tracks;
an 8\% error in the lowest $Q^2$ bin due to the calorimeter energy scale; and a
9\% (20\%) error on the lower four bins (highest $Q^2$ bin)
due to uncertainties in the hadronic energy correction.
A radiative correction factor, $\delta r_{CC}$,
 in the range 1.02 to 1.03 has been applied
to the visible cross section in each bin in order to report a
Born cross section.

\section{Conclusions}

\begin{figure}[tbp]
\epsfxsize=6in
\epsfysize=6.5in
\epsfbox[-50 40 517 607]{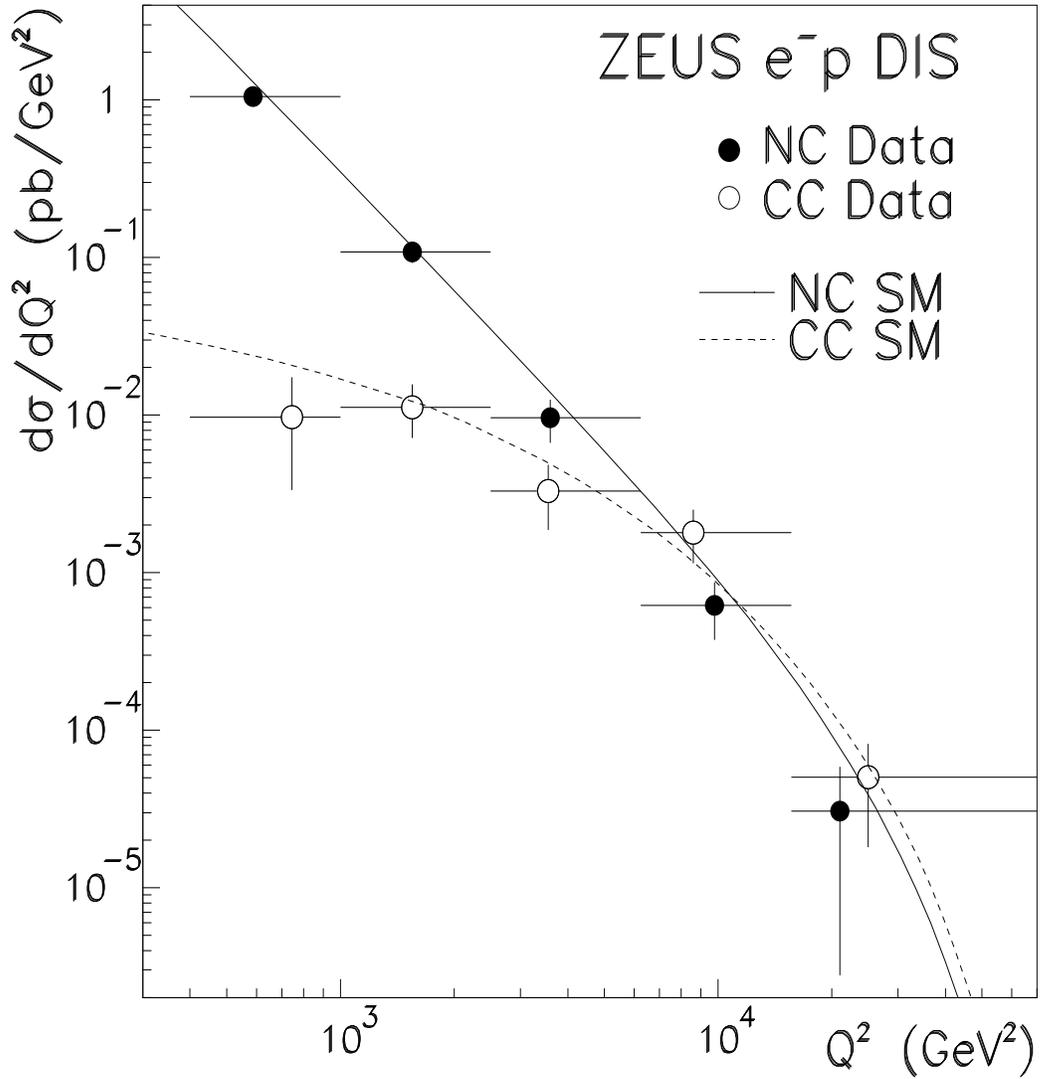}
\vspace*{22mm}
\caption[fig5]{\label{fig:dsdq} $d\sigma/dQ^2$ for
CC and NC DIS. The points with errors are the data, and the
curves are the Standard Model cross sections.
The data are plotted at the average $Q^2$ of the events in each bin.}
\end{figure}

The differential Born cross sections $d\sigma/dQ^2 $ for both NC and CC
 scattering are shown in
Figure \ref{fig:dsdq}.
The measured cross sections agree  with the Standard Model predictions.
The ratios of the NC to CC total cross sections for $Q^2 > Q^2_{min}$ are
listed in Table \ref{tab:dndbin}.
{}~From the lowest bin in $Q^2$ to the highest (for which $Q^2 \simeq M_W^2$),
the ratio of $d\sigma_{NC}/dQ^2 $ to $d\sigma_{CC}/dQ^2 $
decreases by two orders of magnitude to around unity,
thus demonstrating the equal strengths
of  the weak and electromagnetic forces at high $Q^2$.

The rapid fall of the NC cross section with increasing
$Q^2$ is mainly due to the massless photon
propagator, as can be seen from the contribution to the NC cross section
from photon exchange only, $\sigma^{\gamma}_{NC}$ in Table \ref{tab:dndbin}.
The $Q^2$-dependence of the CC cross section is  sensitive to the
mass, $M_W$,  in the CC propagator.
The CC cross sections expected in the limit
of infinite propagator mass, $\sigma^{M_W\rightarrow\infty}_{CC}$,
are inconsistent with the data, as shown
in Table \ref{tab:dndbin}.
Fitting $d\sigma_{CC}/dQ^2$ with $M_{W}$ as the free parameter, and
$G_F$ fixed,
we find
$M_{W} = 76\pm 16(stat)\pm 13(syst) \mbox{\rm ~GeV}$, which
agrees with  the $W^{\pm}$ mass,
 $M_W=80.22\pm0.26$~GeV \cite{PDG}, measured at hadron colliders.

\section{Acknowledgements}

The ZEUS experiment has
been made possible by the ingenuity and dedicated effort of many people
from inside DESY and from the outside institutes who are not listed here
as authors.
Their contributions are deeply appreciated, as are
the inventiveness and continued diligent efforts of the HERA machine group
and the DESY network computing services.
We thank the DESY directorate for strong support and encouragement.

%

\newpage
\vspace{3cm}
\begin{table}[ht]
\vspace{3cm}
\begin{tabular}{|l| c c c c c |}\hline
$Q^2_{min}, Q^2_{max} (\mbox{\rm ~GeV}^2) $
& $ ~~400, ~1000 $ & $ ~1000,~2500 $ &
 $ ~2500,~6250  $ & $  ~6250,15625  $ & $15625, 87500$ \\ \hline
$N_{NC}$  & 328 & 86 & 18 & 3 & 1 \\
$\sigma^{meas}_{NC}(pb) $  & 629 $\pm38\pm69$ & 163 $\pm18\pm15$ &
36 $\pm9\pm4$ & 5.8 $^{+3.6}_{-3.2} \pm 0.6$ & 2.0 $^{+2.9}_{-1.6}\pm 0.3$\\
$\delta r_{NC}$ & 0.89 & 0.88 & 0.89 & 0.91 & 0.95 \\
$\sigma^{SM}_{NC} (pb) $  & 644 & 167 & 41 & 8.8 & 1.1 \\
$\sigma^{\gamma}_{NC}(pb)$   & 636 & 159 & 35 & 5.9 & 0.6 \\ \hline
$N_{CC}$ & 2 & 7 & 5 & 7 & 2\\
$\sigma^{meas}_{CC}(pb)$  & 5.8  $^{+4.6}_{-3.8}\pm 0.9$ &
 16.8 $^{+6.7}_{-6.1}\pm 2.3$  & 12.3 $^{+5.8}_{-5.3}\pm 1.7$ &
 16.8 $^{+6.7}_{-6.1}\pm 2.2$ & 3.4 $^{+2.7}_{-2.1} \pm 0.8$ \\
$\delta r_{CC}$ & 1.02 & 1.03 & 1.03 & 1.03 & 1.02 \\
$\sigma^{SM}_{CC} (pb) $ & 13.3 & 17.1 & 15.9 & 8.0 & 1.6 \\
$\sigma^{M_W\rightarrow \infty}_{CC}(pb)
 $ & 17.5 & 28.3 & 41.8 & 46.0 & 21.2 \\ \hline
$\sigma_{NC}(Q^2>Q^2_{min}) $ & $837 \pm 100$ & $209 \pm ~27$
& $~46 \pm 12 $  & $~8.0\pm ~4.1$ & $~2.0 \pm 1.7$ \\
$\sigma_{CC}(Q^2>Q^2_{min}) $ & $ ~57 \pm ~20$  & $~50 \pm ~13 $  &
$ ~34 \pm ~10 $  & $~21 \pm ~3.1$ & $~3.4 \pm 2.7$ \\
$R
\left(\frac{\sigma_{NC}}{\sigma_{CC}}\right)_{Q^2>Q^2_{min}}
$ &
$14.7 ^{+3.4}_{-3.2}$ &  $4.2 ^{+1.3}_{-0.9}$ & $ 1.4^{+0.6}_{-0.4}$ &
$0.4^{+0.3}_{-0.1}$ & $0.7^{+1.0}_{-0.5}$ \\ \hline
\end{tabular}

\caption[tab:binrates]{ \label{tab:dndbin} Events observed and
integrated Born cross sections
for NC and CC DIS.
Errors shown are statistical, followed by
systematic (which includes the 3.5\% luminosity uncertainty).
The Born cross sections were obtained from the visible cross sections
by multiplying by the radiative correction factor, $\delta r_{NC,CC}$.
The Standard Model (SM) cross sections are calculated with
LEPTO\cite{lepto} using the
$\mbox{\rm MRSD}^\prime_-$  parton distributions\cite{mrsd}. The predictions
for a photon-only NC $\sigma_{NC}^{\gamma}$, and
for an infinite mass in the CC propagator
$\sigma^{M_W\rightarrow \infty}_{CC}$, are also shown.
The NC to CC cross sections and their ratios, $R$,
are also given for  $Q^2 > Q^2_{min}$.}
\end{table}

\end{document}